\documentclass[reprint,
superscriptaddress,
amsmath,amssymb,
aps,prl,
twocolumn,
showpacs,
]{revtex4-1}
\usepackage{tikz}
\usepackage{graphicx}
\usepackage{dcolumn}
\usepackage{bm}
\usepackage{braket}
\usepackage{hyperref}
\usepackage{amsmath}
\usepackage{graphicx}
\usepackage{fancyhdr}
\usepackage{float}
\usepackage{todonotes}
\usepackage{xcolor}
\usepackage{xspace}
\usepackage{eucal}
\usepackage{ulem}
\usepackage{todonotes}
\usepackage{multirow}
\usepackage[utf8]{inputenc}
\usepackage{pgfplots}
\pgfplotsset{compat=1.14}
\usepackage{upgreek}
\usepackage{siunitx}
\usepackage{gensymb}
\usepackage{textcomp}
\usepackage{comment}
\usepackage{siunitx}
\usepackage{todonotes}
\usepackage{cleveref}
\usepackage{siunitx}
\DeclareSIUnit\g{g}
\DeclareSIUnit\gal{Gal}
\DeclareSIUnit\torr{Torr}
\DeclareSIUnit\inch{inch}
\DeclareSIUnit\joule{J}
\begin{document}
\title{Optomechanical lasers for inertial sensing}
\author{Hayden Wisniewski}
\author{Logan Richardson}
\author{Adam Hines}
\author{Alexandre Laurain}
\author{Felipe~Guzman}\email[Electronic mail: ]{felipe@optics.arizona.edu}
\affiliation{James C. Wyant College of Optical Sciences, University of Arizona, 1630 E. University Blvd., Tucson, AZ 85721, USA}%
\date{\today}
%
\begin{abstract}
We have developed an inertially sensitive optomechanical laser by combining a Vertical-External-Cavity Surface-Emitting Laser with a monolithic fused silica resonator. By placing the external cavity mirror of the VECSEL onto the optomechanical resonator test mass, we create a sensor where external accelerations are directly transcribed onto the lasing frequency. We developed a proof-of-principle laboratory prototype and observe test mass oscillations at the resonance frequency of the sensor through the VECSEL lasing frequency, $\num{4.18}\,\pm\,$\SI{.03}{\hertz}. In addition, we set up an ancillary heterodyne interferometer to track the motion of the mechanical oscillator's test mass, observing a resonance of $4.194\pm0.004\ \mathrm{Hz}$. The interferometer measurements validate the VECSEL results, confirming the feasibility of using optomechanical lasers for inertial sensing.
\end{abstract}
\maketitle
\section{Introduction}
Monolithic optomechanical sensors constructed from low-loss materials result in inertial sensors of exquisite sensitivity to external accelerations~\cite{Guzman2014,OptomehanicalInertial}. Such sensors have a compact footprint allowing them to meet the requirements of field or networked measurements, can be incorporated into quantum hybrid inertial sensors~\cite{QuantumOptomehanicalInertial}, and have even been utilized in the post-correction of atom interferometers~\cite{richardson2019opto}. Previously, such sensors~\cite{Guzman2014} leveraged high sensitivity displacement interferometry to readout changes in test mass position and therefore acceleration. While this has proven to be effective, this approach requires an external laser system and associated optical components, which limits the ability to reduce size, weight and complexity of the instrument topology. 

Utilizing Vertical-External-Cavity Surface-Emitting Laser (VECSEL) technology provides a path for reading out the position of inertailly sensitive test masses and an opportunity to construct inertial sensors whose laser metrology source is directly integrated onto the mechanical oscillator~\cite{GuzmanPatentNIST}. VECSELs constitute a subgroup of semiconductor lasers where the gain medium and one cavity mirror is incorporated into a semiconductor chip, while the other cavity mirror or mirrors are external \cite{ct:kuz97,ct:Kuz99,SDLbook}. VECSEL chips are manufactured with a specific gain bandwidth, however, the peak emission wavelength, $\lambda_e$ is dependent on cavity length. By placing the external cavity mirror of the VECSEL onto an inertially sensitive optomechanical oscillator test mass, we create a sensor where peak emission wavelength is dependent on test mass displacement, and therefore acceleration. 

Changing the peak wavelength over the gain bandwidth has been demonstrated by combining micro-electromechancial systems (MEMS) and Vertical-Cavity Surface-Emitting Lasers (VCSELs) to create tunable lasers \cite{ct:MemsVC,ct:MemsVC2}. This principle of an oscillating end mirror has been previously explored in the context of high frequency cavity optomechanics which utilize high frequency MEMS oscillators ($f_0 \geq \SI{1}{\kilo\hertz}$) mounted with high contrast gratings (HCG)~\cite{ct:OptomechLaser,baldacci2016thermal,ct:OptomechDiss}. While these measurements were aimed at measuring the internal forces caused by radiation pressure and thermal effects, we apply these principles to create a low-frequency inertially sensitive optomechanical laser, with applications in gravimetry and long period accelerometry. 

In this article, we present the development of a proof-of-principle optomechanical laser and functional demonstration measurements that highlight the feasibility of this approach for inertial measurements. The optomechanical laser is a mechanically dynamic system that is simultaneously a fully operating laser comprised of an GaAs based VECSEL with the external mirror mounted on the test mass of a \SI{10}{\hertz} monolithic fused silica mechanical resonator. We set up an ancillary heterodyne interferometer to conduct verification measurements.  Currently, the VECSEL chip is optically pumped from an external \SI{808}{\nano\meter} pump laser diode, although we envision future versions of this technology operating with electronically pumped VECSELs, enabling a self-contained sensor. Our prototype optomechanical laser observes clear mechanical signatures through the VECSEL peak emission output, such as the mechanical oscillator's resonance, thus tracking the motion of the test mass. 

\section{Background}
\subsection{Monolithic resonators}
The mechanical resonator constitutes the inertially sensitive element in the combined optomechancal laser. Inertially sensitive optomechanical resonators, when combined with a readout system, make excellent accelerometers with numerous advantages over commercially available sensors including vacuum compatibility, external field insensitivity and relative compactness. These sensors are micro-fabricated from laser assisted dry etching, using low loss materials, which gives a high degree of design control of the optomechanical resonator. This allows us to tailor the sensor's resonance and geometry to optimize performance for a given application.

\begin{figure}[htbp]
\centering
\fbox{\includegraphics[width=.95\linewidth]{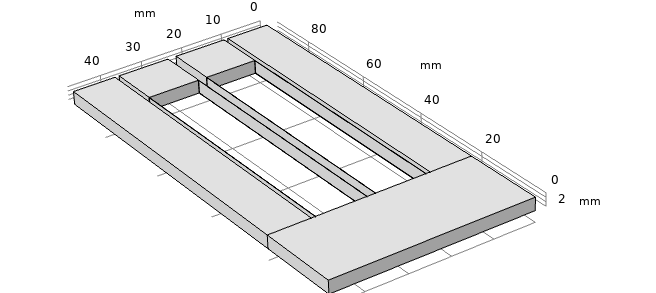}}
\caption{Rendering of the fused silica monolithic resonator. Two \SI{0.95}{\gram} test masses are attached to the U-shaped frame by two $\SI{100}{\micro\meter}$ thick legs. The two thin legs provide rectilinear motion at a designed resonate frequency of $10\ \mathrm{Hz}$. It has overall dimensions of $\SI{48}{\milli\meter} \times \SI{92}{\milli\meter} \times \SI{3}{\milli\meter}$ and mass of \SI{26}{\gram}. The semiconductor chip of the laser is mounted on the frame while the external cavity mirror is mounted on one of the test masses.}
\label{SensorFig}
\end{figure}

The resonator utilized in this work is designed to measure low frequency accelerations, with an unladen resonance of \SI{10}{\hertz}, a footprint of \SI{48}{\milli\meter} $\times$ \SI{92}{\milli\meter} and a total mass of \SI{26}{\gram}. A rendering of this sensor design can be seen in Figure~\ref{SensorFig}. The resonator is comprised of a set of two test masses, each supported by two legs that are $\SI{100}{\micro\meter}$ thick in the direction of motion. Test mass motion is rectilinear provided by the two-legged flexure design. Each test mass is capable of measuring accelerations independently, and the second mass is currently present for redundancy.

Dynamics of the resonator test mass informs how external accelerations drive the test mass displacement, as well as determining the fundamental noise floor of such a sensor. The equation of motion for the optomechanical resonator test mass are given by \cite{ct:TransferFnPaper},

\begin{equation}
    F = m\ddot{x} + m\Gamma_{v}\dot{x} + m\omega_{0}^{2}(1+i\phi(\omega))x
    \label{eqn_of_motion}
\end{equation}
where $m$ is the test mass, $\omega_{0}$ is the resonance frequency, $\Gamma_{v}$ is the velocity damping rate governed by gas damping, and $\phi(\omega)$ is loss coefficient for internal losses. The velocity damping term can be mitigated by operating in vacuum, and is not considered in further discussions in this paper for simplicity. An in-depth performance analysis of this sensor, including loss mechanisms, is discussed in~\cite{OptomehanicalInertial}.

Hence, the test mass displacement linear spectral density can be expressed as,

\begin{equation}
    X^2(\omega)=\frac{4k_{B}Tk\phi(\omega)}{\omega((k-m\omega^2)^2+k^2\phi^2}
    \label{displacementEq}
\end{equation}
where $k$ is the spring constant, $k_B$ is Boltzmann's constant, $T$ is temperature, $\phi$ is the frequency dependent damping and $\omega$ is the angular frequency of oscillation \cite{ct:TransferFnPaper,OptomehanicalInertial}. Using Equation~\ref{displacementEq}, the acceleration spectral density can be calculated with the transfer function,
\begin{equation}
    \frac{X(\omega)}{A(\omega)}=\frac{-1}{\omega^2_0-\omega^2+i\frac{\omega_0}{Q}\omega}
    \label{TransferEq}
\end{equation}
where $A(\omega)$ is the acceleration spectral density, $\omega_0$ is the resonance frequency of the mechanical oscillator, and $Q$ is the mechanical quality factor.

The acceleration resolution of these sensors, $a_{th}(\omega)$, is limited fundamentally by thermal noise, which is described by,

\begin{equation}
    a_{th}(\omega)=\sqrt{\frac{4k_{B}T\omega_{0}}{mQ}},
    \label{accelResEq}
\end{equation}

The smaller $a_{th}$ becomes, the higher the acceleration resolution of the sensor. Maximizing the product of $m$ and $Q$ will maximize the acceleration sensitivity at room temperature. The quality factor is then related to the loss angle by,
\begin{equation}
    Q = \frac{1}{\phi(\omega_0)}
    \label{lossEq}
\end{equation}
which aids in calculating losses in the system. Main sources of loss come from gas damping, bulk, surface, anchor, and thermoelastic losses. A detailed discussion on losses in these resonators is found in~\cite{OptomehanicalInertial}, from which we measured an $mQ$-product of \SI{250}{\kilo\gram}, yielding a projected sensitivity up to $\sim\SI{1e-11}{\meter\per\square\second\per\sqrt{\hertz}}$ \cite{OptomehanicalInertial}.

As previously stated, the mechanical resonators are designed to have a resonance of \SI{10}{\hertz}, however, additional mass is introduced into the mechanical system by the VECSEL's external cavity mirror mounted onto the oscillating test mass, resulting in a lower resonance frequency. In our case, we added to mirrors onto the test mass: a)~an external VECSEL cavity mirror with a mass of $\SI{0.6}{\gram}$, and b)~a small silvered mirror for the heterodyne laser interferometer with a mass of $\SI{0.48}{\gram}$. The total added mass of approximately $\SI{1.1}{\gram}$ lowered the mechanical resonance to $\omega_0 \approx \SI{4}{\hertz}$.

At lower frequencies, Equation~\ref{TransferEq} can be approximated by ${d}a\approx \omega^2 {d}x$. In our laboratory setup the system is not isolated, and therefore exposed to external accelerations, acoustics and vibrations on the optical table on the order of $\widetilde{a}\approx\SI{1e-4}{\meter\per\second\squared}$, which correspond to a test mass displacement on the order of $~\num{1}-\SI{5}{\micro\meter}$.

\subsection{{Vertical-External-Cavity Surface-Emitting Lasers}}
 VECSELs are semiconductor lasers that are a variant of the more commonly constructed Vertical-Cavity Surface-Emitting Laser (VCSEL). VCSELs are semiconductor lasers where the cavity and the gain medium, two of the components needed to make a laser, are combined on a semiconductor chip \cite{ct:VCSEL1,ct:VCSEL2,ct:VCSEL3}. A gain medium is grown between two different distributed Bragg reflectors (DBRs), which is a set of quarter wavelength thick layers of high and low index material designed for high optical reflectively \cite{ct:DBR}. One of the DBRs also acts as an output coupler for the laser. A VECSEL has only the DBR behind the gain medium on the chip \cite{ct:kuz97,ct:Kuz99,SDLbook}. The second VCSEL DBR, which acts as the output coupler, is replaced with an external mirror in a VECSEL. 

VECSELs provide several advantages over other laser technologies, including large wavelength versatility from $\SI{440}{\nano\meter}$ to $\SI{5}{\micro\meter}$~\cite{UVVEC,MidIRVEC}, large wavelength tunable range by actuating the cavity length, and high optical output powers when optically pumped. The VECSEL chip used in this prototype sensor has a designed nominal wavelength at $\Bar{\lambda}=\SI{1042}{\nano\meter}$. In the case of this work, the laser has demonstrated wavelength emission in the range of $\lambda_{e}=\num{1042}-\SI{1046}{\nano\meter}$. 

VECSELs can either be pumped electrically or optically \cite{SDLbook}.  Optical pumping is achieved through absorption of a pump laser in the active region of the semiconductor chip. Due to the large broadband absorption of the semiconductor material, this method is simpler to implement as growing the chip is less intensive. Furthermore, the ability to move the position of the pump introduces another degree of freedom when aligning the system. This also allows us to pump the VECSEL with high optical power, which allows for high VECSEL outputs of the order of tens of Watts. However, a drawback of optically pumped VECSELs is that it requires the external laser pump and external optics to focus the pump beam onto the chip, resulting in larger systems. Electrically pumping a VECSEL requires that the chip be manufactured with electrodes that access the p-n junction that forms from the active region. An electrical current is then applied across this junction, which creates light in the gain medium. Electrically pumping has drawbacks in the form of efficiency degradation due to intracavity laser absorption and non uniformity of injection over a large aperture \cite{SDLbook}, resulting in much lower VECSEL output powers. However, this latter point is not a concern for our low power optomechanical laser. Our current proof-of-principle system is optically pumped, however, we envision the development of future iterations of this technology to be electrically pumped, allowing for compact stand-alone devices.

A change in cavity length, $\Delta L$, corresponds to changes in frequency as,

 \begin{equation}
    \Delta \nu = \frac{c}{\Bar{\lambda}}\frac{\Delta L}{L},
     \label{freqEq}
 \end{equation}
where $\Delta \nu$ is the change in frequency, $L$ is the cavity length, $c$ is the speed of light, and $\bar{\lambda}$ is the average VECSEL wavelength.

Readout of the VECSEL output frequency change under test mass dynamics can be achieved by beating the output beam against a stable reference laser, or by measuring the VECSEL frequency directly using an spectrometer or a wavemeter.

\section{Sensor concept}
\begin{figure}[htbp]
\centering
\fbox{\includegraphics[width=.95\linewidth]{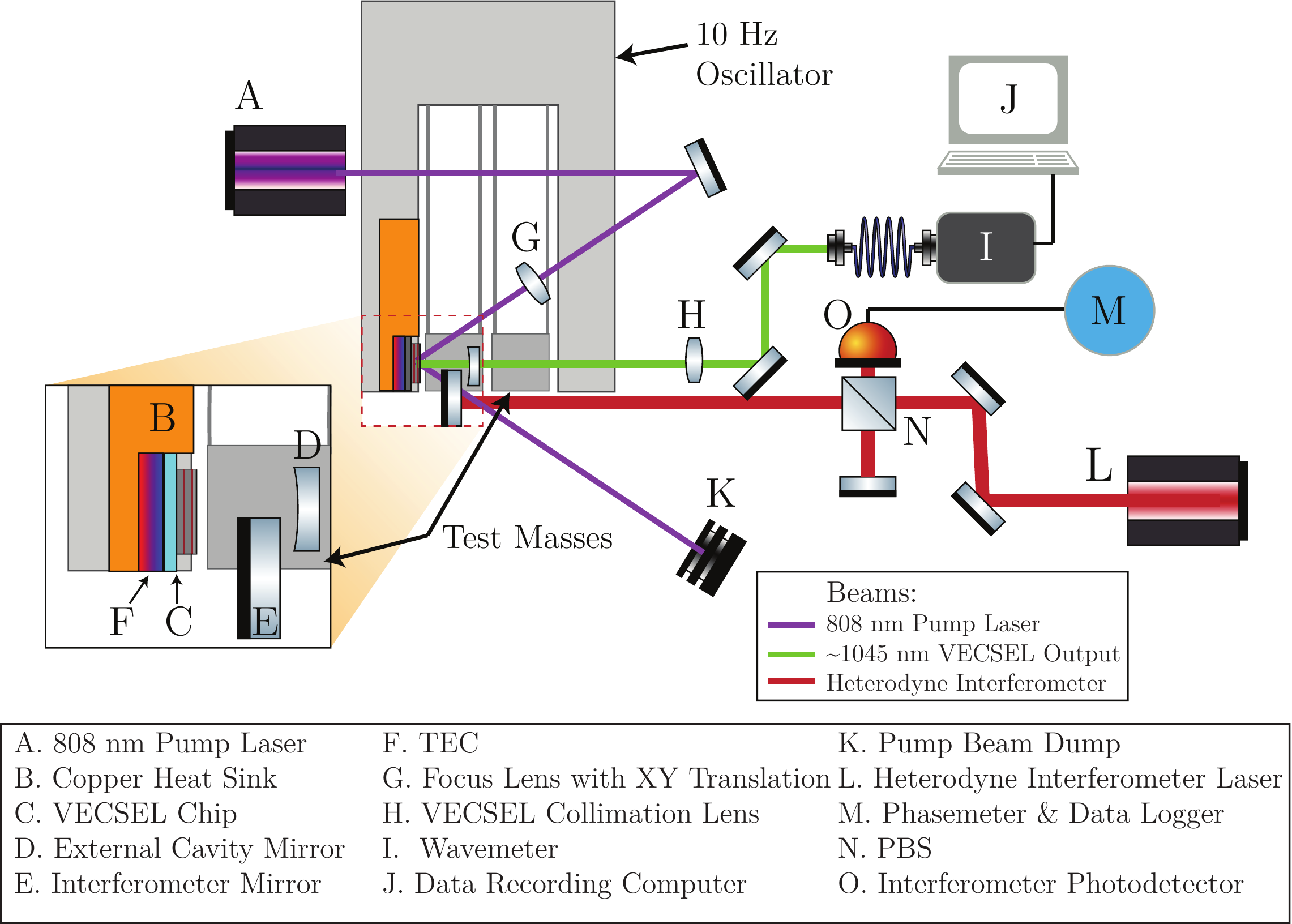}}
\caption{Optomechanical laser (B-D, F) being collimated and focused into the wavemeter (H-J). Laser mirror (D) is attached to the oscillating test mass of the resonator, which induces peak wavelength oscillation when moving. Mirror E is also attached to the mass and provides secondary motion data through the ancillary heterodyne Mach-Zehnder interferometer(E, M-O), which uses a Zygo 7701A HeNe-laser (L) and a polarizing beam splitter (N).}
\label{GOLdiagramFig}
\end{figure}

Here we describe the design and construction of the inertially sensitive optomechanical laser composed of the fused silica resonator, the VECSEL chip, a thermoelectric cooler (TEC), and a copper heat sink. Figure~\ref{GOLdiagramFig} shows a diagram of the setup.

The pump and its focusing system are external optics which are not intrinsically part of the optomechanical sensor. An L-shaped aluminum mount was constructed to allow for repeatable placement of the mechanical oscillator in the same position without the use of a clamping system. The resonator frame rests on a U-shaped piece of glass that matches the frame of the resonator, which lifts the test masses from the surface and allows them to oscillate freely. The current system has been set up on an optical table and is operated in air without additional mechanical or environmental isolation.

The VECSEL incorporated here uses a resonant periodic gain multi-quantum well structure with InGaAs quantum wells and GaAs barriers for the active region. The chip has a surface area of $\SI{4}{\milli\meter} \times \SI{4}{\milli\meter}$ with a thickness of a few microns. This active region contains the gain medium, as well as the DBR mirror to form one end of the laser cavity. The DBR mirror is bonded to a diamond chip for heat management. This particular VECSEL chip was grown by the University of Marburg, Germany, in collaboration with the University of Arizona~\cite{ct:chipgrowth}.

\begin{figure}[htbp]
\centering
\fbox{\includegraphics[width=.95\linewidth]{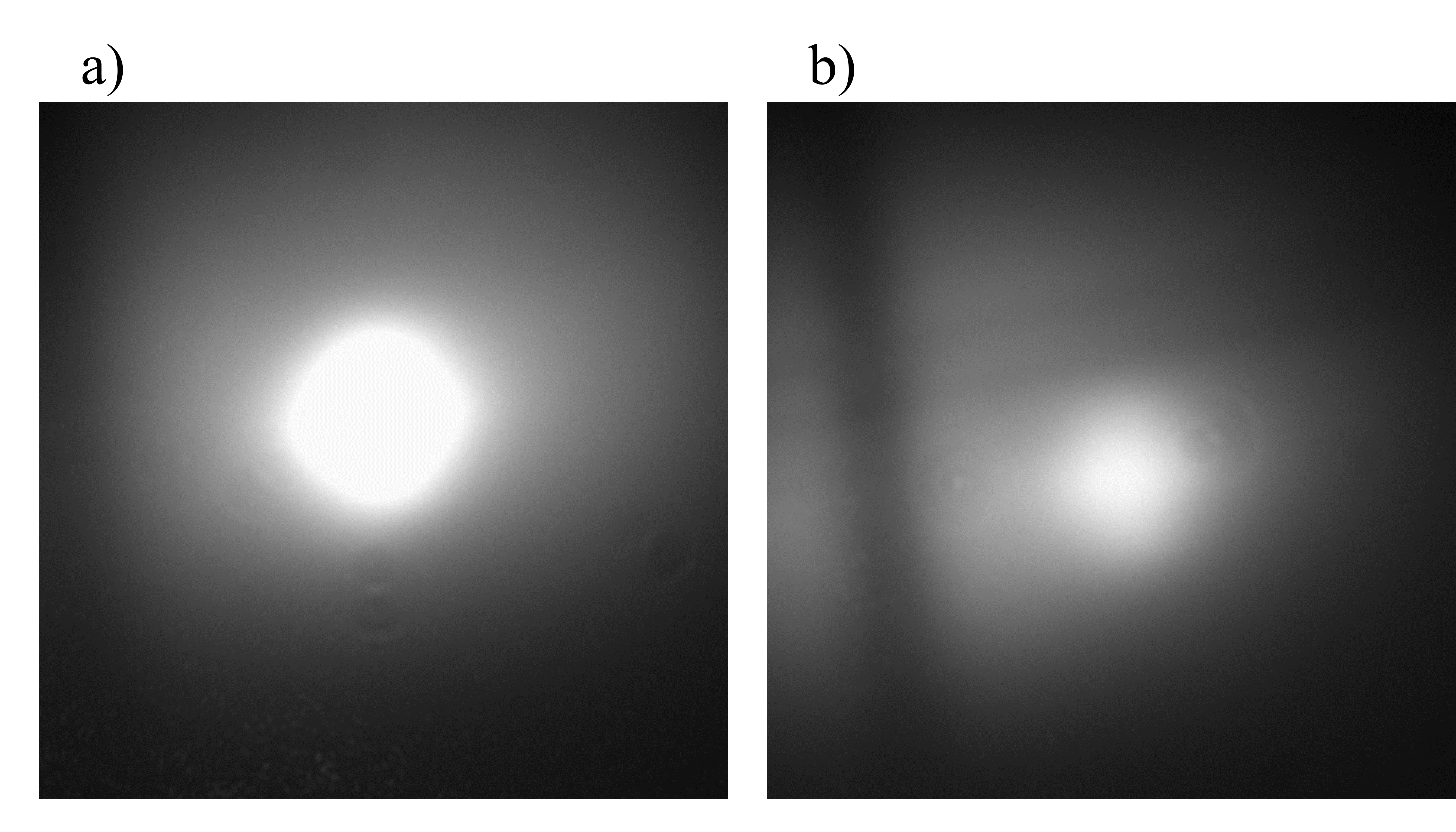}}
\caption{Pump beam as seen from the surface of the VECSEL chip imaged using a CCD.	a) The \SI{808}{\nano\meter} laser pump hitting a defect free portion of the chip surface. For this sensor, the minimum spot diameter to induce lasing was \SI{140}{\micro\meter}. b) The laser pump hitting a chip surface that has manufacturing defects, seen here as three dark lines. The mirror and focusing lens for the pump were used to direct the beam away from such defects.}
\label{fig:vecselsurface}
\end{figure}

The VECSEL pumping optics consist of an \SI{808}{\nano\meter} pump laser, a mirror in a kinematic mount, and a singlet lens in an X-Y translation mount. The pump laser is directed by the steering mirror onto the chip at an angle of $\phi = \ang{35}$ with respect to the surface normal. The chip surface normal is the optical axis of the cavity and, therefore, the lasing direction. Light is focused from the pump beam onto the chip using a $f=\SI{50}{\milli\meter}$ singlet lens, producing a spot radius of \SI{70}{\micro\meter}. Imaged by a CCD (Flir Chameleon 2.0) and using the degrees of freedom provided by the mirror and lens movement, the pump beam is adjusted until it impinges upon a spot with no surface defects on the chip, as shown in Figure~\ref{fig:vecselsurface}. The lasing threshold of the VECSEL was determined to have a pump density of \SI{1}{\kilo\watt\per\centi\meter\squared}, corresponding to a pump output power of \SI{153.9}{\milli\watt}.

The DBR reflector of the VECSEL chip forms one half of a Fabry-Perot laser cavity, with the other half formed by a custom manufactured Altechna mirror with a radius of curvature $R=\SI{50}{\milli\meter}$ and a reflectivity of $r<99.5\%$, which is placed onto the inertial sensitive test mass. The system will continue to lase as long as it stays within the stable cavity regime and therefore the external mirror may move along the optical axis as far as \SI{50}{\milli\meter} away from the chip surface. We aligned the external cavity mirror of the VECSEL and placed it onto the test mass using a three-axis translation stage and a kinematic mount, allowing for control of 4 degrees of freedom - 3 translation axes and one rotation. Once we aligned the mirror and achieved lasing, we secured the VECSEL external mirrorusing a UV-curing epoxy adhesive. After the curing time, we released the mirror from the alignment jig while maintaining a lasing VECSEL system.

We collimate the output VECSEL beam using an {$f=\SI{100}{\milli\meter}$} singlet lens. We then couple this into an optical fiber. The highest VECSEL power we measured in the optical fiber was \SI{3.5}{\milli\watt}.

Thermal fluctuations are in the same low frequency band as the accelerations we intend to measure with our optomechanical laser and are a major noise source considered in this system. To further reduce the effect of thermal fluctuations, we perform thermal stabilization using a temperature controlled TEC that is thermally bonded to the VECSEL chip. Heat is transferred from the TEC to a copper heat sink that is bonded to the frame of the resonator. The copper sink provides not only a thermal sink, but it is also the mechanical interface of the VECSEL to the optomechanical oscillator. Therefore, a change in the thickness of the heatsink would couple into to the cavity length, and ultimately into the inertial measurement. Thus, the thermally driven cavity length change is given by: $\Delta \ell = \alpha \ell \Delta T$, where $\Delta \ell$ is the change in thickness, $\ell$ is thickness at nominal temperature, $\Delta T$ is the change in temperature, and $\alpha$ is the coefficient of thermal expansion - $\SI{16.8e-6}{\meter\per\meter\per\kelvin}$ for copper \cite{Mechbook}. We then set the TEC to maintain a chip temperature of {\SI{293.15}{\kelvin}}.

We machined the portion of the heatsink directly behind the chip and TEC to a thickness of \SI{1}{\milli\meter}. This minimizes the thermally sensitive portion of the sink that can affect the cavity length, while still being able to efficiently transfer heat from the chip. Assuming a \SI{5}{\kelvin} temperature fluctuation, much larger than what was ever observed, the \SI{1}{\milli\meter} thick heat sink would correspond to a \SI{84}{\nano\meter} cavity length change. This is negligible when compared to the \SI{1}{\micro\meter} cavity length change expected from non-isolated accelerations experienced by the mechanical oscillator in the laboratory. 

The current prototype has a cavity length of $\SI{9}{\milli\meter}$ and an average wavelength of $\bar{\lambda} =\SI{1042}{\nano\meter}$. Non-isolated laboratory accelerations are on the order $\Tilde{a}\approx\SI{1e-4}{\meter\per\second\squared}$. Using Equations~\ref{TransferEq} and \ref{freqEq}, accelerations at this level correspond to test mass displacements on the order of \SI{4.6}{\micro\meter} from equilibrium and to a change in emission wavelength of the optomechanical laser of $\Delta \lambda_e = \SI{0.5}{\nano\meter}$, or $140\,\mathrm{GHz}$ laser frequency change. This change in wavelength is difficult to measure through a standard beatnote measurement, and hence we utilized a wavemeter (Burleigh WA-1100) to measure to output wavelength (see Figure~\ref{GOLdiagramFig}).

An independent measurement system utilizing a heterodyne laser interferometer (Zygo 7701A) was constructed to simultaneously track the motion of the test mass displacement and corroborate the measurements taken by the spectrometer and the wavemeter.  The HeNe-laser used produced two overlapping, perpendicularly polarized beams of light separated by a frequency of \SI{25}{\mega\hertz}. A polarizing beam-splitter separates one polarization of the light to a reference arm and the other to a mirror which is mounted on the same test mass as the external VECSEL mirror. We recombined the light from the two arms at the beam-splitter and sent it to a photodetector which measured the phase difference in the interfering light produced by the change in optical path length. We measured the phase of the interferometer light using a phasemeter (Liquid instruments Moku:Lab). 

\section{Results}
We recorded measurements of the test mass dynamics using the wavemeter and the reference heterodyne interferometer. We mechanically excited the test mass from equilibrium causing the resonator to naturally oscillate at its resonant frequency. We fiber coupled emission of the VECSEL into a wavemeter and the VECSEL wavelength was converted to displacement using Equation~\ref{freqEq}. Simultaneously, we recorded the phase of the heterodyne interferometer using the phasemeter, and converted it to displacement by: $\Delta L = {(4\pi)^{-1}}{\bar{\lambda} \Phi}$, where $\Phi$ is the measured phase of the interferometer. 

\begin{figure}[htbp]
\centering
\fbox{\includegraphics[width=.95\linewidth]{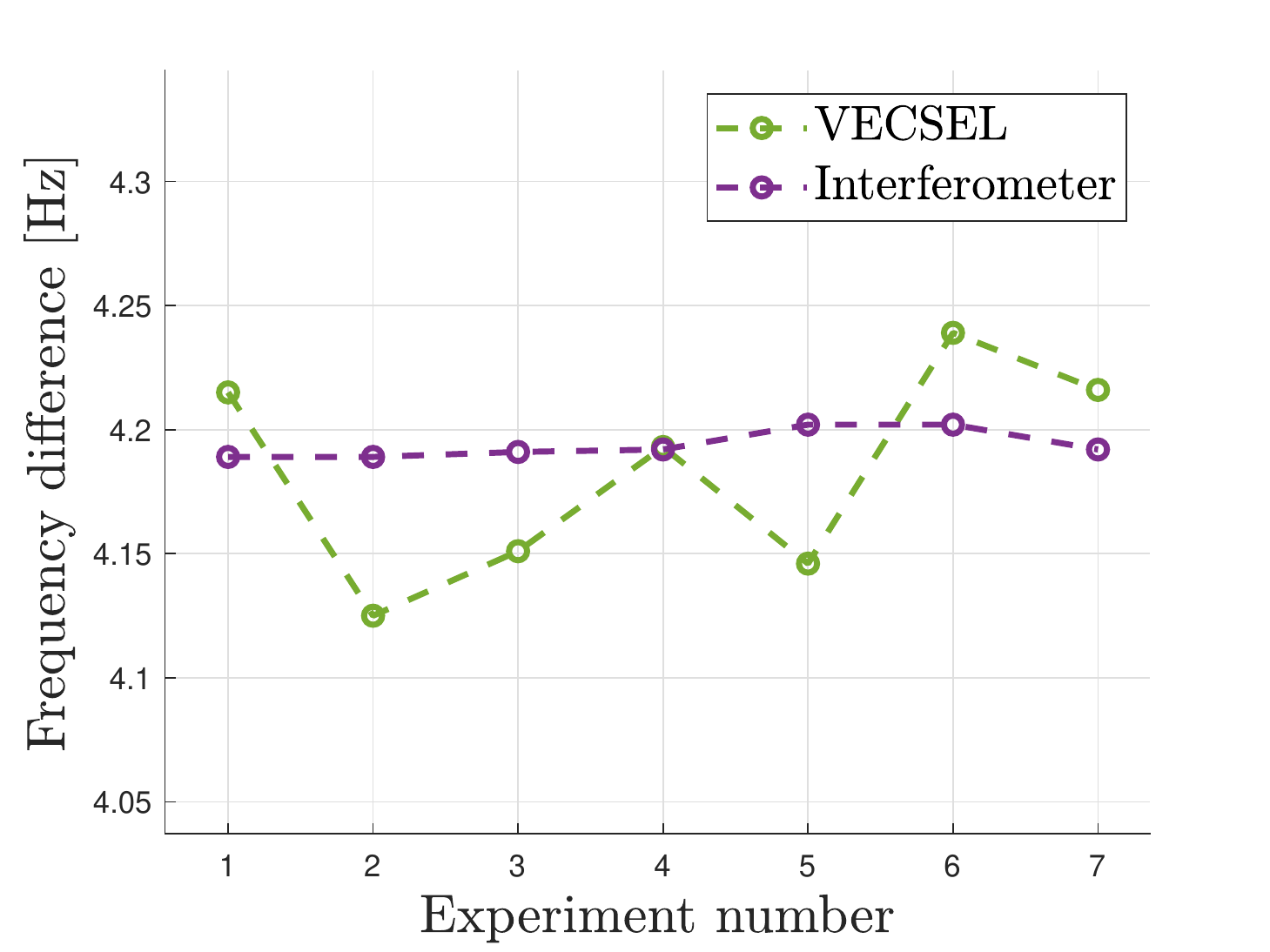}}
\caption{Frequency measured by the VECSEL and the interferometer is plotted along the left axis. The right axis shows the difference between the two values for each trial. The VECSEL measured an average frequency of $4.18\pm 0.03\ \mathrm{Hz}$, and the interferometer measured an average frequency of $4.194\pm0.004\ \mathrm{Hz}$.}
\label{fig:freqDiff}
\end{figure}

We performed each measurement run over a duration of 30 minutes, from which we computed the primary oscillation frequency for each experiment. Due to the high inertial noise level in the laboratory during these experiments, we first focused on determining the mechanical resonant frequency using both methods, the optomechanical laser and the heterodyne interferometer. The resonance frequency measurement results are shown in Figure~\ref{fig:freqDiff}. The VECSEL output as measured by the wavemeter resulted in an average frequency of $4.18\pm 0.03\ \mathrm{Hz}$, whereas the interferometer measured an average frequency of $4.194\pm0.004\ \mathrm{Hz}$.

Emission from the VECSEL periodically exhibited multi-mode behavior, which degraded the broadband measurement resolution of the VECSEL output using the wavemeter. An analysis of the multi-mode behavior is treated further in the Section~\ref{sec:conclusion}.  For a more discerning measurements of the peak emission of the VECSEL, we utilize an spectrometer (ASEQ HR-1) which recorded the full spectral output of the VECSEL from \num{963.08}-\SI{1067.41}{\nano\meter} with a resolution of $\delta \lambda > \SI{.35}{\nano\meter}$ every \SI{14}{\milli\second}. This readout technique, however, is limited to acquiring 50 continued spectrum measurements. 

\begin{figure}[htbp]
\centering
\fbox{\includegraphics[width=.95\linewidth]{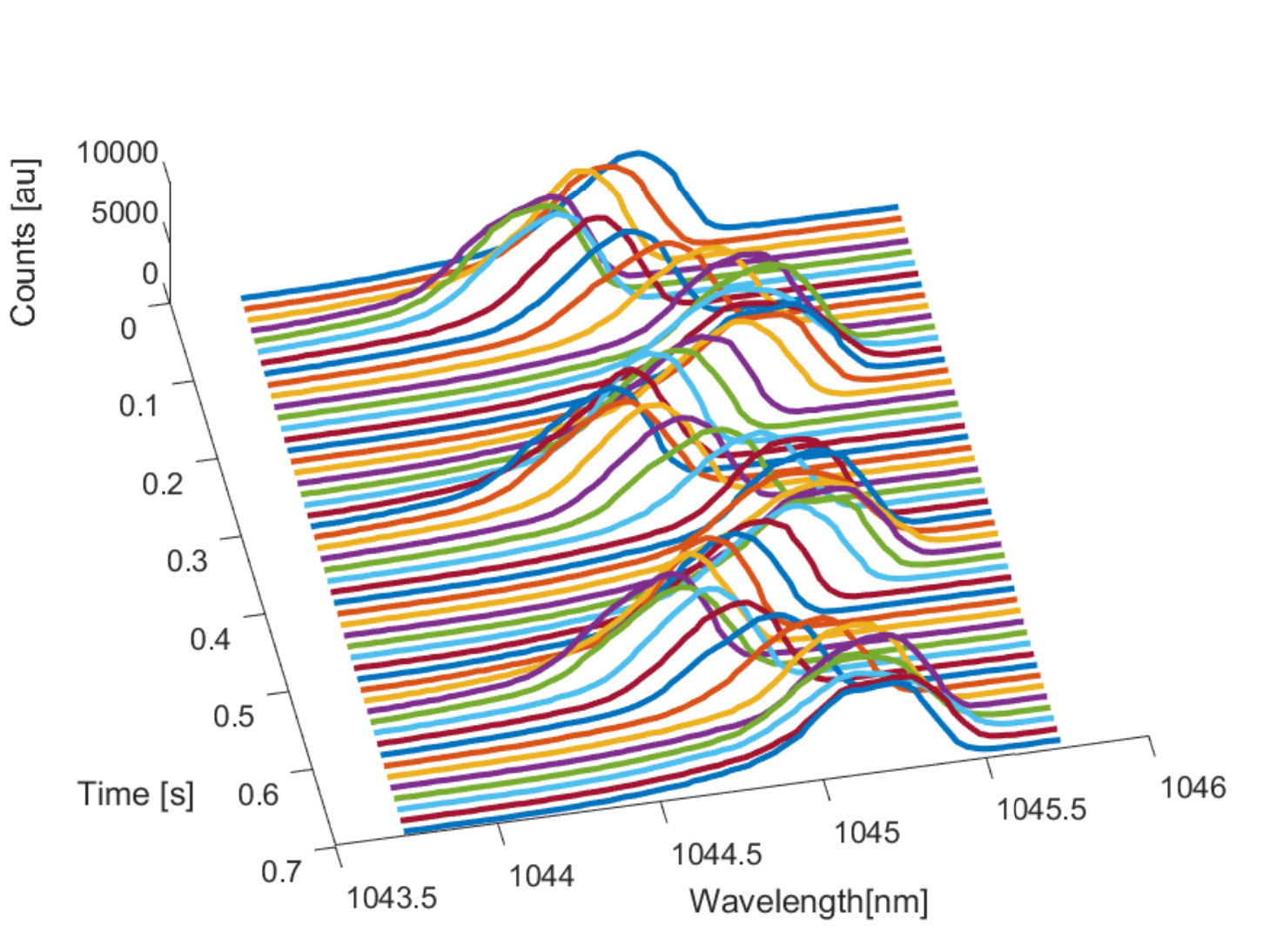}}
\caption{Raw output of 50 spectral scans collected by an ASEQ HR1 spectrometer taken over a \SI{0.7}{\milli\second} period. The oscillation of the full spectrum can be seen as the spectra evolve over the time period. The peak wavelength of each spectrum was taken and compared against the interferometer data in Figure~\ref{fig:displacement}.}
\label{fig:3dosa}
\end{figure}

\begin{figure}[htbp]
\centering
\fbox{\includegraphics[width=.95\linewidth]{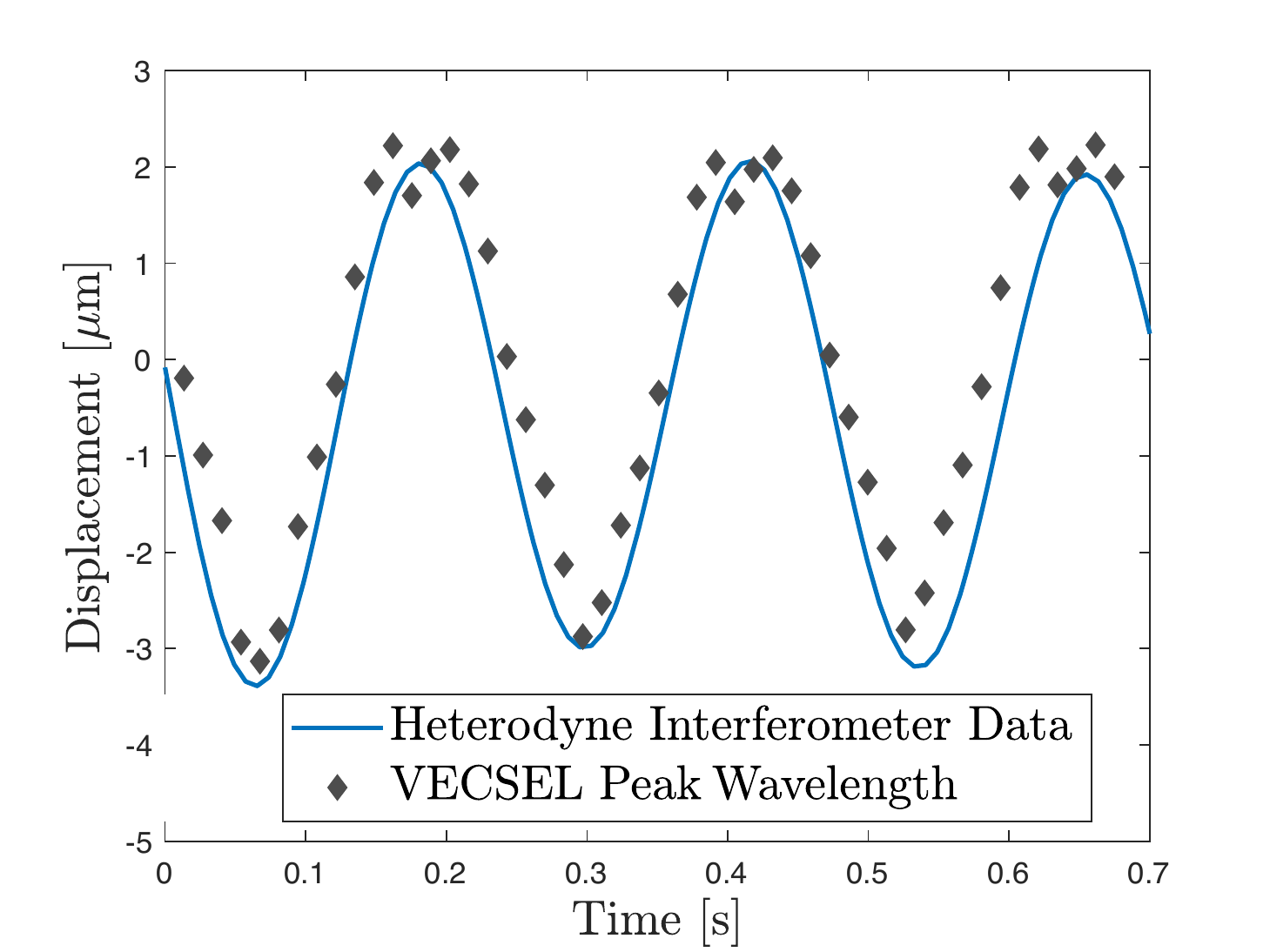}}
\caption{Displacement of the interferometer mirror as measured by the Zygo heterodyne interferometer for a given time interval. We recorded the spectral output of the VECSEL over the same interval by the spectrometer. Each of the 50 spectrometer spectra, spaced \SI{14}{\milli\second} apart, were fitted and the peak wavelength was returned. This data was then converted from change in wavelength to displacement using Equation~\ref{freqEq}, and plotted over the heterodyne interferometer data.}
\label{fig:displacement}
\end{figure}

In order to observe the characteristic test mass dynamics of the oscillator, we introduced a mechanical excitation and recorded the test mass displacement using the spectrometer and the heterodyne interferometer.

A plot showing the series of acquired spectra by the spectrometer during this measurement run can be seen in Figure~\ref{fig:3dosa}. We fit a Gaussian model to each spectra and converted the peak wavelength value to a displacement using Equation~\ref{freqEq}. Figure~\ref{fig:displacement} shows the comparison between the spectrometer and the interferometer measurements when both are converted to displacement. It can be seen that both instruments, our optomechanical laser and the hetordyne interoferometer observe the same test mass displacement, both in displacement amplitude and oscillation period, which validates the output of the optomechanical laser as an inertial sensor.

\section{Discussion}
\label{sec:conclusion}

We demonstrated proof-of-principle measurements of an inertially sensitive optomechanical laser that consists of a protoype VECSEL laser with a dynamic end mirror, which is capable of observing the dynamics of a test mass through changes in the lasing frequency.

This proof-of-principle system currently exhibits higher noise than the heterodyne interferometer due to the unstable measurement environment, and construction of the sensor. Furthermore, the VECSEL output exhibited occasional multi-mode behavior, which we can attribute to the relatively long cavity length required for operation, 9\,mm. A shorter cavity length ($L\leq \SI{5}{\milli\meter})$ would allow a more stable single mode operation~\cite{Laurains}, however, due to the geometry of the VECSEL chip, heat sink, pump laser and external cavity mirror, this was not feasible in this prototype. Additionally, the current in-air operation environment allows for high coupling of acoustics and vibrations.

In addition, operation at atmospheric pressures limits the performance of the optomechanical laser due to gas damping, which limits the mechanical quality factor $Q$. To mitigate this, we are currently preparing the laser for in-vacuum operations. Fused silica resonators of the same design have been measured to have a $Q$ value of above $1\times 10^{5}$ \cite{OptomehanicalInertial}. With larger $Q$ values, the acceleration noise floor of the optomechanical sensor will improve. In addition, other limiting factors that need to be addressed include anchor losses from how the resonator is mounted, resonator design focused on increasing $Q$, and long term VECSEL single-mode output lasing stability. We will address these limiting factors in future developments of optomechanical lasers for inertial sensing.

Moreover, future iterations of this system rely on advancing the technology of electrically pumped VECSELs to ensure reliable and stable long term operations, and enable fully integrated and self-contained optomechanical lasers. Lastly, achieving highly sensitive measurements with optomechanical lasers outlines a path for optical frequency-referenced inertial sensing as a primary reference metrology technology.

\section{Funding Information}
This work was supported, in part, by the National Geospatial-Intelligence Agency (Grant Number: HMA04761912009) and the National Science Foundation (Grant Number: PHY-1912106).

\section{Acknowledgments}
The authors would like to acknowledge Mitul Dey Chowdhury for work on the ancillary interferometer, Andrea Nelson for assistance configuring the wavemeter for data collection, and Robert Rockmore for stimulating discussions. The authors thank Cristina Guzman for reviewing and improving parts of the manuscript.

\section{Disclosures}
The authors declare no conflicts of interest.

\bibliography{references}

\end{document}